\documentclass[epsfig,psfig,aps,twocolumn,prb,showpacs]{revtex4-1}
\usepackage{amsfonts}
\usepackage{graphicx}
\usepackage{epsfig}
\usepackage{epsf}

\begin{document}
\draft
\title
{
Long-wavelength optical phonon behavior in uniaxial strained graphene: Role of electron-phonon interaction\\ 
}
\author
{
Mohamed Assili and Sonia Haddad 
} 
\address{
Laboratoire de Physique de la Mati\`ere Condens\'ee, D\'epartement de Physique,
Facult\'e des Sciences de Tunis, Universit\'e Tunis El Manar, Campus Universitaire 1060 Tunis, Tunisia
}
%\maketitle
%\date{\today}
%
%---------Abstract---------
%
\begin{abstract}
We derive the frequency shifts and the broadening of $\Gamma$ point longitudinal optical (LO) and transverse optical (TO)
phonon modes, due to electron-phonon interaction, in graphene under uniaxial strain as a function of the electron density 
and the disorder amount. We show that, in the absence of a shear strain component, such interaction gives rise to a lifting 
of the degeneracy of the LO and TO modes which contributes to the splitting of the G Raman band. The anisotropy of the electronic 
spectrum, induced by the strain, results in a polarization dependence of the LO and TO modes. This dependence is in agreement with 
the experimental results showing a periodic modulation of the Raman intensity of the splitted G peak.
Moreover, the anomalous behavior of the frequency shift reported in undeformed graphene is found to be robust under strain.
\end{abstract}

\maketitle

\section{Introduction}
Since its discovery in 2004 \cite{Geim}, graphene continues to be the subject of intense interest regarding its exotic properties
\cite{revcastroneto,revmark}. These intriguing properties, such as the anomalous quantum Hall effect, are ascribed to Dirac type 
electrons described by the Weyl's equation for massless particles \cite{revmark}. The electronic properties in graphene are significantly 
affected by applying a strain \cite{Pucci}. The latter can also, accidentally, occur during the fabrication process as in exfoliation 
or chemical vapor deposition of graphene samples \cite{Peres}. \newline
Theoretical and first principle calculations revealed the substantial effect of the strain on the electronic and lattice spectra 
of graphene \cite{Mohr1,Cheng,Vozmed1,Vozmed2}.\

To bring out the signature of strain induced modified electronic and vibrational properties, Raman spectroscopy has emerged as a powerfull probe.
This technique, which is simple to use in graphene, is found to be a successful tool to identify the number of layers in multilayer graphene, 
to probe the nature of disorder and the doping amount \cite{Dresselhaus1,Dresselhaus2,Basko}.\

Several experimental studies have been carried out on Raman spectra of graphene under uniaxial strain 
\cite{Ferralis2008,Ninano,MHuang,NiPRB,Mohiuddin,frank0,frank1,frank2,Lee2012,C-huang2013}. 
The results revealed that, due to the strain, the Raman G band is redshifted and splitted into two peaks denoted G$^+$ and G$^-$. G$^+$ (G$^-$) 
is the mode polarized perpendicular (along) the strain direction. The G peak appearing in unstrained graphene at 1580 cm$^{-1}$ corresponds 
to a doubly degenerate optical mode at the $\Gamma$ point of the Brillouin zone (BZ). The splitting of the G peak results from the strain induced 
lattice symmetry lowering.\

Experimental results showed that the frequency shift rates of the G$^+$ and G$^-$ as a function of the strain 
strength $\epsilon$ is of $\frac{\partial \omega_{G^-}}{\partial\epsilon}\sim$ -13 cm$^{-1}\,/\%$ 
and $\frac{\partial \omega_{G^+}}{\partial\epsilon}\sim$ -6 cm$^{-1}\,/\%$ \cite{MHuang}.
Recent measurements \cite{C-huang2013,Mohiuddin,frank2,Son2011,MinHuang} reported that the rate 
shifts of G$^-$ and $G^+$ are respectively of -33 cm$^{-1}\,/\%$ and -14 cm$^{-1}\,/\%$ in agreement with first principle 
calculations \cite{Mohiuddin,Son2011}. The difference in the shift rates was attributed to strain calibration \cite{Son2011}.
The G band splitting could be understood within a phenomenological model based on a semiclassical approach \cite{Mohiuddin,Popov,Thomsen}. 
Within this model, the shear component of the strain is found to be responsable of the splitting.\

Raman spectroscopy of strained graphene has also revealed that the 2D band, originating from a resonant scattering 
process involving two optical phonons at the BZ edges, splits into two peaks under uniaxial strain \cite{MinHuang,Mohr,Son2011}. 
This splitting was ascribed to strain induced changes in the resonant conditions resulting from both modified electronic band 
structure and phonon dispersion \cite{Son2011,Popov}.\

Several studies reported that the electron-phonon coupling plays a key role in Raman spectroscopy in graphene 
\cite{castro2007,revcastroneto,Sasaki2012,Yan07}. Ando \cite{ando2006} showed that, in undeformed graphene, 
the frequency of the center zone optical phonon mode is shifted  due to electron-phonon interaction. The frequency 
behavior is found to depend on the value of the Fermi energy $E_F$ compared to the phonon frequency $\omega_0$ at the $\Gamma$ point:
For $E_F<\frac {\hbar \omega_0} 2$ ($E_F>\frac {\hbar \omega_0} 2$), the phonon frequency is redshifted (blueshifted) leading 
to a lattice softening (hardening).
In the clean limit, a logarithmic singularity takes place at $E_F=\frac {\hbar \omega_0} 2$ which is found to be smeared out 
in the dirty limit and at finite temperature \cite{Lazzeri}.
Moreover, Ando\cite{ando2006} reported an anomalous behavior of the optical phonon damping induced by the electron-phonon 
interaction: for $E_F<\frac {\hbar \omega_0}2$, the phonons are damped due to the formation of electron-hole pairs leading 
to phonon softening \cite{revcastroneto}. However, for $E_F>\frac {\hbar \omega_0}2$, the phonon is no more damped since the 
electron-hole pair production is forbidden by Pauli principle \cite{ando2006,revcastroneto}. This damping behavior predicted 
by Ando \cite{ando2006} was observed in Raman spectroscopy \cite{Pisana07,Yan07,revcastroneto}.\

The natural question, which arises at this point, is how the frequency shifts and damping of optical phonon are modified in 
uniaxial strained graphene where electron band structure is deeply changed.\

Theoretical studies \cite{Pereira,mark2008} showed that the perfect honeycomb lattice of graphene undergoes a quinoid-type 
deformation by applying a uniaxial strain. The Dirac cones are no longer at the corners of the BZ and are tilted.
The corresponding low energy electronic properties could be described by the generalized tow dimensional (2D) 
Weyl's Hamiltonian \cite{mark2008}. It is worth to note that the tilted Dirac cones are also expected in the organic 
conductor $\alpha$-(BEDT)$_2$I$_3$ where BEDT stands for bis(ethylenedithio)-tetrathiafulvalene \cite{suzumura,suzumura2,morinari,mark2008}.\
Based on the generalized Weyl's Hamiltonian, several intriguing properties of this compound have been unveiled \cite{suzumura,morinari,mark2008,assili}.\

In this paper, we focus on the effect of the electron-phonon interaction on the $\Gamma$ point optical phonon modes 
in graphene under uniaxial strain described by a quinoid-type lattice. We show that the frequency shift and the broadening 
of the longitudinal optical (LO) and the transverse optical (TO) phonon modes are substantially dependent on the characteristic 
parameters of the Weyl Hamiltonian which are the tilt and the anisotropy of the electronic dispersion relation. 
We bring out original points which, to the best of our knowledge, have not been addressed so far: (i) the electron-phonon 
interaction in strained graphene induces a lifting of the degeneracy of the LO and TO modes which contributes to the splitting of the G band. 
This effect is found to originate from the anisotropy of the electronic spectrum and not from the tilt of Dirac cones. The latter may only give 
rise to a global shift of the G band compared to the undeformed case. The splitting is found to be strongly dependent on the electron 
density and disorder amount.
(ii) The anomalous behavior of the phonon damping reported in Refs.\cite{ando2006, revcastroneto} in undeformed graphene is found to be
a robust feature which is kept under uniaxial strain. The damping of LO and TO modes strongly depends on the strain amplitude and the phonon angle. 
We found that, in the particular case, where one of the mode is along the strain direction, the corresponding phonons are strongly damped for 
a compressive deformation. However the phonon mode perpendicular to the strain direction is less damped and its lifetime increases as the strain 
amplitude increases. For tensile deformation the mode behaviors are exchanged.
(iii) A crossing of TO and LO frequencies can take place at a particular doping values as found in carbon nanotubes \cite{Sasaki2008}.
(iv) We found that the electron-phonon interaction contributes to the polarization dependence of the G peak in uniaxial strained graphene 
as concluded by Mohiuddin {\it et al.}\cite{Mohiuddin}.\

The paper is organized as follows: In Sec. II we give the outlines of the formulation to derive the optical phonon self-energy. We start with 
the generalized Weyl's Hamiltonian obtained within the effective mass approach. Then, we derive the electron-phonon interaction Hamiltonian 
and the phonon self-energy. The results are discussed in Sec. III in relation with experiments. Sec. IV is devoted to the concluding remarks.

\section{Optical phonon self-energy}
We consider the optical phonon modes of the center BZ responsable of the G peak in graphene. We focus on the LO and inplane TO modes.
We first derive the electronic Hamiltonian, within the effective mass theory \cite{divencinzo,revando,Macucci}, taking into account the 
first and second neighbor hopping parameters in strained graphene.
\subsection{Electronic Hamiltonian}
\begin{figure}[hpbt] 
\begin{center}
\includegraphics[width=0.7\columnwidth]{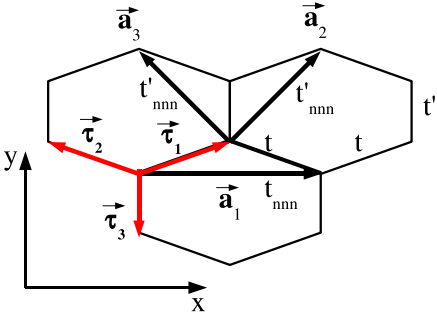}
\end{center}
\caption{Deformed honeycomb lattice along the $y$ axis. ($\vec{a}_1,\vec{a}_2$) is the lattice basis. The hopping parameters to the 
first (second) neighbors $t$ and $t^{\prime}$ ($t_{nnn}$ and $t_{nnn}^{\prime}$) are different due the deformation. Vectors connecting 
first (second) neighboring atoms are denoted $\vec{\tau}_l$ ($\vec{a}_l$).}
\label{Figlattice}
\end{figure}

By applying a uniaxial strain along, for example, the $y$ direction the honeycomb lattice turns to a quinoid type lattice \cite{revmark}. 
It is worth to note that one should consider an arbitrary strain direction as done for example in Refs.\cite{Mohr1,Pereira}. However, 
several experimental and numerical studies \cite{Mohiuddin,Mohr1} have shown that the G band behavior is independent of the strain direction. 
Considering a generic strain direction will give rise to the same form of the electronic Hamiltonian but with renormalized parameters. 
We, then, consider for simplicity a strain along the $y$ direction as in Ref.\cite{revmark}. 
In such case, the hopping parameter to the first neighboring atoms are no more equal as in undeformed graphene. The distance between 
neighboring atoms along the $y$ direction changes from $a$ to 
\[
a^{\prime}=a+\delta a\]
The vectors, $\vec{\tau}_l$ ($l=1,2,3$), connecting the sites of the A sublattice with first neighbors sites on the B sublattice are given 
by (Fig.\ref{Figlattice}):
\begin{eqnarray}
\vec{\tau}_1&=&\frac a 2 \left( \sqrt{3}\vec{e}_x+\vec{e}_y\right),\;
 \vec{\tau}_2=\frac a 2 \left( -\sqrt{3}\vec{e}_x+\vec{e}_y\right),\;\nonumber\\
\vec{\tau}_3&=&-a(1+\epsilon)\vec{e}_y 
\end{eqnarray}
where $a$ is the distance between first neighbor atoms in undeformed graphene,  $\epsilon=\frac{\delta a} a$ is the lattice deformation 
which measures the strain amplitude. $\epsilon$ is negative (positive) for compressive (tensile) deformation.\\

The second neighbors sites are connected by vectors $\vec{a}_l$ given by:
\begin{eqnarray}
\vec{a}_1&=&\sqrt{3}a\vec{e}_x,\,
\vec{a}_2=\frac{\sqrt{3}}2a\vec{e}_x+a\left(\frac32 +\epsilon\right)\vec{e}_y,\nonumber\\
\vec{a}_3&=&-\frac{\sqrt{3}}2a\vec{e}_x+a\left(\frac32 +\epsilon\right)\vec{e}_y 
\end{eqnarray}
where ($\vec{a}_1,\vec{a}_2$) is the lattice basis.\

The hopping integral along $\vec{\tau}_3$ is affected by the strain and is different from those along $\vec{\tau}_1$ and $\vec{\tau}_2$ 
which are equal. Moreover, the hopping parameters to the second neighboring atoms along $\vec{a}_2$ and $\vec{a}_3$ are modified by the 
strain compared to that along $\vec{a}_1$.\

It is worth to stress that by applying a strain along the $y$ direction one should expect a strain component along the $x$ axis $\epsilon_{xx}=-\nu \epsilon_{yy}$ where $\nu=0.165$ is the Poisson ratio of graphene. The off diagonal terms of the strain tensor, which depend on the strain direction and the Poisson ratio\cite{Pereira}, generate different bond lengths. However, for a strain axis parallel to the principal symmetry direction $x$ or $y$, these terms vanish leading to equal bond lengths as assumed in our model. The contribution of Poisson ration could, then, be neglected compared to the main contribution resulting from the strain component along the stress axis.\

%%%%%%%%%%%Figure%%%%%%%%%%%%%%%%%
%
We denote by $t_{nn}^{(l)}$ ($t_{nnn}^{(l)}$) the hopping integral to the first (second) neighboring atoms along $ \vec{\tau}_l$ ($\vec{a}_l$) vectors.
We set $t_{nn}^{(1)}=t_{nn}^{(2)}=t$.\
Under strain $t_{nn}^{(3)}$ changes from $t$ to $t^{\prime}$ given by \cite{mark2008} 
\[
t^{\prime}=t+\frac{\partial t}{\partial a}\delta a 
\]
$t_{nnn}^{(l)}$ along $\vec{a}_2$ and $\vec{a}_3$ changes from the value of undeformed graphene, denoted $t_{nnn}$, to $t^{\prime}_{nnn}$ written as:
\[
t^{\prime}_{nnn}=t_{nnn}+\frac{\partial t_{nnn}}{\partial a}\delta a 
\]
The momentum vectors of Dirac points $D$ and $D^{\prime}$ are given respectively by \cite{mark2008}
\begin{eqnarray}
k^D_y=0,\; \; k^D_x=\xi \frac 2{\sqrt{3}a}\arccos\left(-\frac {t^{\prime}}{2t}\right) 
\end{eqnarray}
where $\xi=\pm$ is the valley index. We denote hereafter 
\begin{eqnarray}
\theta=\arccos\left(-\frac {t^{\prime}}{2t}\right)
\label{theta} 
\end{eqnarray}
In undeformed graphene, the Dirac points $D$ and $D^{\prime}$ are at the corners of the BZ $K$ and $K^{\prime}$. Under the strain, $D$ and $D^{\prime}$ 
move away from $K$ and $K^{\prime}$ points \cite{Pereira,gilles2009}.\\

The electronic wave function can be written as \cite{ando2006,Macucci}:
\begin{eqnarray}
 \psi(\vec{r})=\sum_{\vec{R}_A}\psi_A(\vec{R}_A) \varphi(\vec{r}-\vec{R}_A)+
\sum_{\vec{R}_B}\psi_B(\vec{R}_B) \varphi(\vec{r}-\vec{R}_B)\nonumber\\
\end{eqnarray}
where $\varphi(\vec{r}-\vec{R}_A)$ and $\varphi(\vec{r}-\vec{R}_B)$ are atomic orbitals centred on atoms A and B respectively. \\

In the $\vec{k}.\vec{p}$ approach \cite{revando,Macucci}, the coefficients $\psi_A(\vec{R}_A)$ and $ \psi_B(\vec{R}_B)$ are given by:
\begin{eqnarray}
\psi_A(\vec{R}_A)=\mathrm{e}^{i\vec{k}^D.\vec{R}_A } F_A^D(\vec{R}_A)+
\mathrm{e}^{i\vec{k}^{D^{\prime}}.\vec{R}_A } F_A^{D^{\prime}}(\vec{R}_A)\nonumber\\
\psi_B(\vec{R}_B)=\mathrm{e}^{i\vec{k}^D.\vec{R}_B } F_B^D(\vec{R}_B)-
\mathrm{e}^{i\vec{k}^{D^{\prime}}.\vec{R}_B } F_B^{D^{\prime}}(\vec{R}_B)
\end{eqnarray}
where $F_A^D, \,F_A^{D^{\prime}},\, F_B^D$ and $F_B^{D^{\prime}}$ are slowly varying envelope functions.\\

Considering second neighbor hopping integrals, the electronic energy obeys to:
\begin{eqnarray}
 \varepsilon \psi_A(\vec{R}_A)=-\sum_{l=1}^3 t_{nn}^{(l)} \psi_B(\vec{R}_A-\vec{\tau}_l)
-\sum_{l=1}^6 t_{nnn}^{(l)} \psi_A(\vec{R}_A-\vec{a}_l)\nonumber\\
\varepsilon \psi_B(\vec{R}_B)=-\sum_{l=1}^3 t_{nn}^{(l)} \psi_A(\vec{R}_B+\vec{\tau}_l)
-\sum_{l=1}^6 t_{nnn}^{(l)} \psi_B(\vec{R}_B-\vec{a}_l)\nonumber\\
\label{eigen}
\end{eqnarray}
where $\vec{a}_4=-\vec{a}_1$, $\vec{a}_5=-\vec{a}_2$ and $\vec{a}_6=-\vec{a}_3$.\\

Within the $\vec{k}.\vec{p}$ method, Eq.\ref{eigen} becomes:

\begin{eqnarray}
 \varepsilon\left(
\begin{array}{c}
F_A^D(\vec{r})\\
F_B^D(\vec{r})
\end{array}
\right)=\left(
 \begin{array}{cc} 
w_{0x}k_x& w_xk_x-iw_yk_y\\
w_xk_x+iw_yk_y & w_{0x}k_x\\
 \end{array}
\right)
\left(
\begin{array}{c}
 F_A^D(\vec{r})\\
 F_B^D(\vec{r})
\end{array}
\right)\nonumber\\
\label{eigen1}
\end{eqnarray}
where $\vec{k}=(k_x,k_y)$ is the wave vector and
\begin{eqnarray}
w_x&=&\sqrt{3}a t \sin \theta, \; w_y=\frac 32 t^{\prime}a(1+\frac 23 \epsilon)\nonumber\\
w_{0x}&=&2\sqrt{3}a(t_{nnn}\sin 2\theta +t_{nnn}^{\prime}\sin \theta) 
\end{eqnarray}
Details of the calculations are given in Appendix A.\\

From Eq.\ref{eigen} we recover the so-called minimal form of the generalized Weyl Hamiltonian \cite{revmark,suzumura2}:
\begin{eqnarray}
 H_{\xi}(\vec{k})=\xi\left( \vec{w}_0.\vec{k}\sigma^0+w_xk_x\sigma^x\right)+w_yk_y\sigma^y
\label{Helec}
\end{eqnarray}
where $\vec{w}_0=(w_{0x},w_{0y}=0)$, $\sigma^0={1\!\!1}$, $\sigma^x$ and $\sigma^y$ are the 2x2 Pauli matrices.
The corresponding dispersion relation is of the form:
\begin{eqnarray}
 \varepsilon_{\lambda}(\vec{k})=\vec{w}_0.\vec{k}+\lambda \sqrt{w_x^2k_x^2+w_y^2k_y^2}
\label{dispers}
\end{eqnarray}
$\vec{w}_0$ is responsable of the tilt of Dirac cones away from the $z$ axis. This term obeys to the condition \cite{mark2008}
\begin{eqnarray}
 \tilde{w}_0=\sqrt{\left(\frac{w_{0x}}{w_x}\right)^2+\left(\frac{w_{0y}}{w_y}\right)^2}<1
\end{eqnarray}
which insures the presence of two energy bands: a positive energy for $\lambda=+$ and a negative energy band for $\lambda=-$ \cite{mark2008}. 
In deformed graphene and for $w_{0y}=0$, $\tilde{w}_0\sim 0.6 \epsilon$ \cite{mark2008}.\

The eigenfunctions of the Hamiltonian given by Eq.\ref{Helec} are of the form:
\begin{eqnarray}
 F(\vec{k},\vec{r})=\frac 1{\sqrt{2S^{\prime}}} \left(
\begin{array}{c}
 1\\
\eta  \mathrm{e}^{i\Phi_{\vec{k}}}
\end{array}
\right)
\mathrm{e}^{i\vec{k}.\vec{r}}
\label{eigenfunc}
\end{eqnarray}
where $\eta=\lambda\xi$ is the chirality index, $S^{\prime}$ is the lattice surface under strain and $\tan \Phi_{\vec{k}}=\frac{w_yk_y}{w_xk_x}$.

\subsection{Electron-phonon interaction}
In this section, we derive the effective Hamiltonian describing the effect of the lattice vibrations on the electronic Hamiltonian. 
Such effect arises from the change of the hopping integrals due to the lattice distortion. This Hamiltonian was obtained by Ando \cite{ando2006} 
in the case of undeformed graphene. We shall determine the electron-phonon interaction Hamiltonian in quinoid-type deformed graphene.\\

The phonon Hamiltonian can be written as \cite{ando2006}
\begin{eqnarray}
H_{ph}=\sum_{\vec{q},\mu}\hbar\omega_{0,\mu}\left(b^{\dagger}_{\vec{q},\mu}b_{\vec{q},\mu}+\frac12\right)
\end{eqnarray}
where $b^{\dagger}_{\vec{q},\mu}$ ($b_{\vec{q},\mu}$) is the creation (annihilation) operator of phonon with wave vector $\vec{q}=(q_x,q_y)$ 
and mode $\mu=$ LO, TO. $\omega_{0,\mu}$ is the $\mu$ mode phonon frequency at the $\Gamma$ point.\\

The relative displacement of the two sublattices A and B in the continuum limit is
\begin{eqnarray}
\vec{u}(\vec{r})=\frac 1 {\sqrt{2}}\left(\vec{u}_A(\vec{r})-\vec{u}_B(\vec{r})\right),
\end{eqnarray}
which can be written for optical phonon at $\Gamma$ point as \cite{ando2006}:
\begin{eqnarray}
\vec{u}(\vec{r})=\sqrt{\frac{\hbar}{2NM}}\sum_{\vec{q},\mu}\frac 1{\omega_{0,\mu}}\left( b_{\vec{q},\mu}+b^{\dagger}_{-\vec{q},\mu}\right)\vec{e}_{\mu}(\vec{q})\mathrm{e}^{i\vec{q}.\vec{r}}
\label{displa}
\end{eqnarray}
where $M$ is the mass of the carbon atom, $N$ is the number of unit cells and $\vec{e}_{\mu}(\vec{q})$ is given by:
\begin{eqnarray}
\vec{e}_L(\vec{q})&=&i(\cos\varphi(\vec{q}),\sin\varphi(\vec{q}))\nonumber\\
\vec{e}_T(\vec{q})&=&i(-\sin\varphi(\vec{q}),\cos\varphi(\vec{q}))
\end{eqnarray}
with $\tan\varphi(\vec{q})=\frac{q_y}{q_x}$.\\

To derive the electron-phonon effective Hamiltonian, we shall determine the effect of the lattice displacement on the hopping integrals.\\

The hopping parameter between first neighboring atoms located at $\vec{R}_A$ and $\vec{R}_A-\vec{\tau}_l$ is changed from $t_{nn}^{(l)}$ to \cite{Ishikawa}:
\begin{eqnarray}
t_{nn}^{(l)}+\frac{\partial t_{nn}^{(l)}}{\partial d_l} \left[|\vec{\tau}_l+\vec{u}_A(\vec{R}_A)-\vec{u}_B(\vec{R}_A-\vec{\tau}_l)|-d_l\right] 
\label{tnn}
\end{eqnarray}
with $d_l=|\vec{\tau}_l|$, $d_1=d_2=a$ and $d_3=a(1+\epsilon)$.
The hopping integral between second neighboring atoms changes from $t_{nnn}^{(l)}$ to
\begin{eqnarray}
t_{nnn}^{(l)}+\frac{\partial t_{nnn}^{(l)}}{\partial a_l} \left[|\vec{a}_l+\vec{u}_A(\vec{R}_A)-\vec{u}_A(\vec{R}_A-\vec{a}_l)|-a_l\right] 
\label{tnnn}
\end{eqnarray}
However, the correction to $t_{nnn}^{(l)}$ terms vanishes for $\Gamma$ point optical phonon modes ($\vec{q}=\vec{0}$).\\

Since the amplitude of the lattice displacement is small compared to the lattice parameter, Eq.\ref{tnn} becomes:
\begin{eqnarray}
t_{nn}^{(l)}+\frac{\partial t_{nn}^{(l)}}{\partial d_l} \frac{\vec{\tau}_l}{d_l}.\left[\vec{u}_A(\vec{R}_A)-\vec{u}_B(\vec{R}_A-\vec{\tau}_l)\right]
\label{tnncorrect}
\end{eqnarray}
In the continuum limit, $\vec{u}_A(\vec{R}_A)-\vec{u}_B(\vec{R}_A-\vec{\tau}_l)\simeq \vec{u}_A(\vec{r})-\vec{u}_B(\vec{r}_A-\vec{\tau}_l)\sim \sqrt{2}\vec{u}(\vec{r}) $.\

The correction to the hopping integrals due to lattice distortion, given by Eq.\ref{tnncorrect}, leads to an extra term $\Delta H$ 
in the electronic Hamiltonian which is written near the D point as (for details, see Appendix B):
\begin{widetext}
\begin{eqnarray}
\Delta H=\frac{\sqrt{2}}{ta}\frac{\partial t}{\partial a} \left(
\begin{array} {cc}
0& w_y^{\prime}u_y(\vec{r})+iw_xu_x(\vec{r})\\
w_y^{\prime}u_y(\vec{r})-iw_xu_x(\vec{r})& 0
\end{array}
\right)\nonumber\\
\label{DeltaH}
\end{eqnarray} 
\end{widetext}
where $u_x(\vec{r})$ and $u_y(\vec{r})$ are the component of the relative displacement $\vec{u}$. $w_y^{\prime}$ is of the form:
\begin{eqnarray}
w_y^{\prime}=-a\left[t\cos\theta +t^{\prime}(1+\epsilon)\right]\sim w_y-2\epsilon t^{\prime} a (1+\epsilon)
\end{eqnarray}
and $\theta$ obeys to Eq.\ref{theta}.\\

Given the expression of $w_y$ and since $t^{\prime}=t(1-2\epsilon)$ \cite{mark2008}, we have $w_y=\frac 32 a t^{\prime}(1+\frac 23 \epsilon)$ and 
$w_y^{\prime}=w_y-\Delta w_y$ with $\Delta w_y=\frac 43 \epsilon w_y$.\\

The electron-phonon Hamiltonian can, then, be written as \cite{ando2006}:
\begin{eqnarray}
H_{int}=-\sqrt{\frac{\hbar}{NM}}\frac{\beta}{a^2}\sum_{\vec{q},\mu}\frac
1{\sqrt{\omega_{0,\mu}}}V_{\mu}(\vec{q})\mathrm{e}^{i\vec{q}.\vec{r}}\left(b_{\vec{q},\mu}+b^{\dagger}_{-\vec{q},\mu}\right)\nonumber\\
\end{eqnarray}
where $\beta=-\frac{d\ln t}{d\ln a}=-\frac a t \frac{\partial t}{\partial a}$, $\omega_{0\mu}$ is the frequency of he optical phonon at $\Gamma$ 
point in the deformed graphene for the mode $\mu$ in the absence of electron-phonon interaction.\

In undeformed graphene $\omega_{0T}=\omega_{0L}=\omega_{0}$.
This degeneracy is expected to be lifted in the strained graphene due to the symmetry breaking. 
According to a phenomenological model \cite{Mohiuddin,Popov,MinHuang} the strain tension $\epsilon_{ij}$ in graphene reduces to $\epsilon_{yy}=\epsilon$ 
where $y$ is the direction of the applied strain, and $\epsilon_{xx}= -\nu \epsilon_{yy}$ along the direction transverse to the strain and $\nu$ 
is the Poisson ratio.
The G band splits into two bands $G^{\pm}$ with frequencies $\omega^{\pm}$ shifted from the unstrained band frequency $ \omega_0$ as
$\Delta \omega^{\pm}= \omega^{\pm}-\omega_0=-\omega_0\gamma_{E_{2g}} (\epsilon_{xx}+ \epsilon_{yy})\pm \frac 12 \beta_{E_{2g}}\omega_0 (\epsilon_{xx}- \epsilon_{yy}) $
where $\gamma_{E_{2g}}$ and $\beta_{E_{2g}}$ are respectively the Gr\"{u}neisen parameter and the shear deformation potential.
The shear component of the strain, $\epsilon_s=\epsilon_{xx}- \epsilon_{yy}$, is then responsible of the G band splitting. 
The question arising at this point concerns the contribution of the electron-phonon interaction to the splitting of the G band.
To highlight this contribution, we did not consider the effect of the shear component which turns out to disregard the effect of the strain 
on the phonon dispersion. We then assume that, in the absence of electron-phonon interaction, the center zone optical phonon modes LO and TO 
have the same frequencies $\omega_{0T}\sim \omega_{0L}\sim \omega_{0}$.
By switching on the interaction, this degeneracy may be lifted giving rise to two bands corresponding to the LO and TO modes which results 
in the G band splitting.\\

The matrices $V_{\mu}(\vec{q})$ are given, near D point, by:
\begin{widetext}
\begin{eqnarray}
V_L(\vec{q})&=&\sqrt{w_xw_y^{\prime}}\left(
\begin{array}{cc}
0& i\frac{\sin \varphi(\vec{q})}{\alpha^{\prime}}-\alpha^{\prime}\cos \varphi(\vec{q})\\
i\frac{\sin \varphi(\vec{q})}{\alpha^{\prime}}+\alpha^{\prime}\cos \varphi(\vec{q})&0\\
\end{array}
\right)\nonumber\\
V_T(\vec{q})&=&\sqrt{w_xw_y^{\prime}}\left(
\begin{array}{cc}
0& i\frac{\cos \varphi(\vec{q})}{\alpha^{\prime}}+\alpha^{\prime}\sin \varphi(\vec{q})\\
i\frac{\cos \varphi(\vec{q})}{\alpha^{\prime}}-\alpha^{\prime}\sin \varphi(\vec{q})&0\\
\end{array}
\right)\nonumber\\
\end{eqnarray}
 \end{widetext}

where $\alpha^{\prime}=\sqrt{w_x/w_y^{\prime}}$. $V_{\mu}(\vec{q})$ near $D^{\prime}$ point 
satisfies $V_{\mu}^{D^{\prime}}(\vec{q})=V_{\mu}^{D}(-\vec{q})^{\ast}$ \cite{ando2006}.\newline
Contrary to acoustic phonons, there is no scalar deformation potential in the interaction Hamiltonian \cite{Suzuura} 
regarding the expression of the relative displacement of the long wavelength optical phonons (Eq.\ref{displa}).
\subsection{Optical phonon self-energy}

The retarded phonon Green function can be written as \cite{ando2006}
\begin{eqnarray}
D_{\mu}(\vec{q},\omega)=\frac{2\hbar\omega_0}{(\hbar\omega+i\eta)^2-(\hbar\omega_0)^2-2\hbar\omega_0\Pi_{\mu}(\vec{q},\omega)}
\end{eqnarray}
$\Pi_{\mu}(\vec{q},\omega)$ is the self-energy and $\eta=\frac{\hbar}{\tau}$, $\tau$ being the scattering time.\\

The shift $\Delta \omega=\omega-\omega_0$ of the phonon frequency is given by the real part of the Green function's pole. For small 
correction to $\omega_0$, $\Delta \omega$ is given by:
\begin{eqnarray}
\Delta\omega=\frac 1{\hbar} \Re\,\Pi_{\mu}(\vec{q},\omega_0)
\end{eqnarray}
The imaginary part of the Green function's pole gives the broadening $\Gamma_{\mu}\propto \frac 1{\tau_{\mu}}$ of the phonon mode. 
$\tau_{\mu}$ being the phonon lifetime:
\begin{eqnarray}
\Gamma_{\mu}=-\frac 1{\hbar} \Im\,\Pi_{\mu}(\vec{q},\omega_0)
\end{eqnarray}

The self-energy of $\Gamma$ point optical phonon can be written as \cite{ando2006,Ishikawa}
\begin{widetext}
\begin{eqnarray}
\Pi_{\mu}(\vec{q}\rightarrow \vec{0},\omega)=-g_vg_s 
\frac{\hbar S^{\prime}}{NM\omega_0}
\left(\frac{\beta}{a^2}\right)^2\sum_{\lambda,\lambda^{\prime}}\int
\frac{d\vec{k}}{(2\pi)^2}
|\langle \lambda^{\prime},\vec{k}|V_{\mu}(\vec{q})|\lambda,\vec{k}\rangle|^2
\frac{f\left(\varepsilon_{\lambda}(\vec{k})\right)-
f\left(\varepsilon_{\lambda^{\prime}}(\vec{k})\right)}
{\hbar \omega+\varepsilon_{\lambda^{\prime}}(\vec{k})
-\varepsilon_{\lambda}(\vec{k})+i\eta}
\label{self}
\end{eqnarray}
\end{widetext}
where $g_v$ and $g_s$ are the valley and spin degeneracy, $f(\varepsilon)$ is the Fermi distribution 
function $f(\varepsilon)=\frac 1{\mathrm{e}^{\frac{\varepsilon-\mu_c}{kT}}+1}$ and $\mu_c$ is the chemical potential at temperature $T$. 
$S^{\prime}$ is the graphene surface under uniaxial strain $S^{\prime}=N \|\vec{a}_1\times \vec{a}_2\| \simeq S\left(1+\frac 23 \epsilon\right) $ 
where $S$ is the undeformed graphene surface.\

For long wavelength phonon modes near $D$ point, the matrix elements can be written as:
\begin{widetext}
\begin{eqnarray}
|\langle \lambda^{\prime},\vec{k}|V_{L}(\vec{q})|\lambda,\vec{k}+\vec{q}\rangle|^2=
\frac{w_xw_y^{\prime}}2\left[\frac{\sin^2\varphi(\vec{q})}{\alpha^{\prime2}}
\left(1-\cos2\Phi_{\vec{k}}\right)+ \alpha^{\prime2}\cos^2\varphi(\vec{q})\left(1+\cos2\Phi_{\vec{k}}\right)
+\sin2\varphi(\vec{q})\cos2\Phi_{\vec{k}}\right]\nonumber\\
|\langle \lambda^{\prime},\vec{k}|V_{T}(\vec{q})|\lambda,\vec{k}+\vec{q}\rangle|^2=
\frac{w_xw_y^{\prime}}2
\left[\alpha^{\prime2}\sin^2\varphi(\vec{q})\left(1+\cos2\Phi_{\vec{k}}\right)+ \frac{\cos^2
\varphi(\vec{q})}{\alpha^{\prime 2}}\left(1-\cos2\Phi_{\vec{k}}\right)
-\sin2\varphi(\vec{q})\cos2\Phi_{\vec{k}}\right]\nonumber\\
\end{eqnarray}
\end{widetext}

According to Eq.\ref{self} only interband processes ($\lambda^{\prime}=-\lambda$)
contribute the self-energy of $\vec{q}=\vec{0}$ phonon modes.\newline

Regarding the electronic dispersion relation (Eq.\ref{dispers}), the term $\hbar \omega+\varepsilon_{\lambda^{\prime}}(\vec{k})
-\varepsilon_{\lambda}(\vec{k})$, in Eq.\ref{self}, becomes 
\[
\hbar \omega+2\lambda \sqrt{w_x^2k_x^2+w_y^2k_y^2} \]
Setting $q_x=w_xk_x$ and $q_y=w_yk_y$, the integration over $\Phi_{\vec{k}}$ in Eq.\ref{self} vanishes and the expression 
of the self-energy can be reduced to an integration over the energy:
\begin{widetext}
\begin{eqnarray}
\Pi_{\mu}(\vec{q}\rightarrow \vec{0},\omega)=-C_{\mu}\int_0^{\varepsilon_c}\frac{\varepsilon d\varepsilon}{2\pi v_F^{\ast 2}}
\left[f\left(-\varepsilon\right)-
f\left(\varepsilon\right)\right]\left[\frac 1
{\hbar \omega+2\varepsilon+i\eta}-
\frac 1{\hbar \omega-2\varepsilon+i\eta}\right]
\label{self3}
\end{eqnarray}
\end{widetext}
where we used the density of state in quinoid lattice $\rho(\varepsilon)=\frac 1{2\pi v_F^{\ast 2}}|\varepsilon|$ \cite{mark2008}. 
$v_F^{\ast}$ is a renormalized Fermi velocity given by \cite{mark2008,FuchsHU}
\begin{eqnarray}
v_F^{\ast}= \sqrt{w_xw_y}\left(1-\frac 34 \tilde{w}_0^2\right) 
\end{eqnarray}
$\varepsilon_c$ in Eq.\ref{self3} is a cutoff energy corresponding to the limit of validity of the linear electronic dispersion 
given by Eq.\ref{dispers} and the coefficient $C_{\mu}$ is given by:
\begin{eqnarray}
C_L&=&A\left[\frac{\sin^2\varphi(\vec{q})}{\alpha^{\prime 2}}+\alpha^{\prime 2}\cos^2\varphi(\vec{q})\right]\nonumber\\
C_T&=&A\left[\alpha^{\prime 2}\sin^2\varphi(\vec{q})+\frac{\cos^2\varphi(\vec{q})}{\alpha^{\prime 2}}\right]
\end{eqnarray}
and $A$ is a constant written as:
\begin{eqnarray}
A=\frac{g_vg_s}4 \frac{36\sqrt{3}}{\pi}\frac{w_y^{\prime}}{w_y}\frac {S^{\prime}}S\frac{\hbar}{2Ma^2\omega_0}\left(\frac{\beta}2\right)^2 
\equiv C\frac{w_y^{\prime}}{w_y}\frac {S^{\prime}}S
\label{A}
\end{eqnarray}

As mentioned in Ref.\cite{ando2006}, one should substract the contribution of $\omega=0$ modes to avoid double counting of electron contribution.
The self-energy at zero temperature takes, then, the form:
\begin{widetext}
\begin{eqnarray}
\Pi_{L}(\vec{q}\rightarrow \vec{0},\omega)=\frac 1{\left(1-\tilde{w}_0^2\right)^{\frac 32}}\left[\frac{\sin^2\Phi}{\alpha^{\prime 2}}
+\alpha^{\prime 2}\cos^2\Phi\right] 
\left[AE_F^{\ast}-\frac 14 A(\hbar \omega +i\eta)\ln\left(\frac{\hbar\omega+2E_F^{\ast}+i\eta}
{\hbar\omega-2E_F^{\ast}+i\eta}\right)+i\pi\right]\nonumber\\
\Pi_{T}(\vec{q}\rightarrow \vec{0},\omega)=\frac 1{\left(1-\tilde{w}_0^2\right)^{\frac 32}}\left[\alpha^{\prime 2}\sin^2\Phi
+\frac{\cos^2\Phi}{\alpha^{\prime 2}}\right] 
\left[AE_F^{\ast}-\frac 14 A(\hbar \omega +i\eta)\ln\left(\frac{\hbar\omega+2E_F^{\ast}+i\eta}
{\hbar\omega-2E_F^{\ast}+i\eta}\right)+i\pi\right]\nonumber\\
\label{self5}
\end{eqnarray}
\end{widetext}
where we set $\Phi=\varphi(\vec{q})$ and $E_F^{\ast}=\hbar v^{\ast}_Fk_F\simeq E_F(1-\frac{\epsilon}3)$ (see Appendix A), 
with $E_F=v_Fk_F$ being the Fermi energy in undeformed graphene.
Eq.\ref{self5} reduces to that obtained by Ando \cite{ando2006} in undeformed graphene for $\alpha^{\prime}=1$ and $\tilde{w}_0=0$.
\section{Results and discussion}
\begin{figure}[hpbt] 
\begin{center}
\includegraphics[width=0.8\columnwidth]{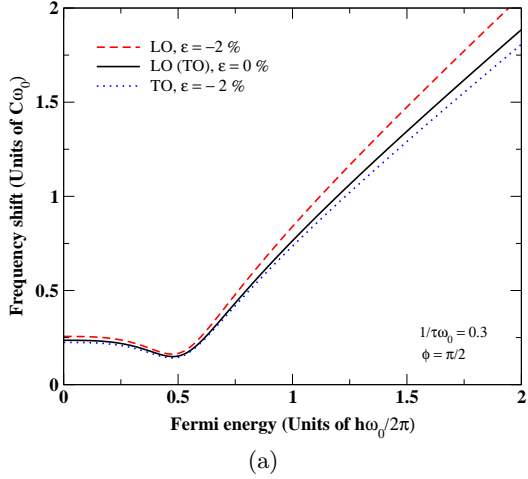}\\

(a)
\vspace{1cm}

\includegraphics[width=0.8\columnwidth]{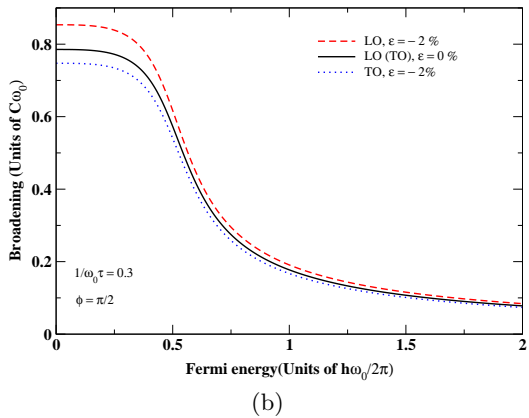}\\

(b)
\end{center}
\caption{Frequency shifts (a) and broadening (b) of LO (dashed line) and TO (dotted line) modes as a function of the 
Fermi energy $E_F$ in the dirty limit $\frac 1{\tau\omega_0}=0.3$ for a compressive strain $\epsilon=\frac{\delta a}a=-2\%$. 
The LO mode is along the strain axis. The solid line is the result for the undeformed case.}
\label{shift02}
\end{figure}

Figures \ref{shift02} shows the dependence of the frequency shifts and broadening of the LO and the TO modes as a function of the 
Fermi energy $E_F$ in the dirty limit for a compressive strain strength $\epsilon=-2\%$. The shifts are normalized 
to $C=A \frac{w_y}{w^{\prime}_y}\frac S{S^{\prime}}$ where $A$ is given by Eq.\ref{A}.\
In undoped system, the effect of electron-phonon interaction on the frequency shifts is not relevant. This effect is 
enhanced by introducing impurities in the system or by increasing the strain amplitude as we will show in the next section.\\

%%%%%%%%Fig.
\begin{figure}[hpbt] 
\begin{center}
\includegraphics[width=0.8\columnwidth]{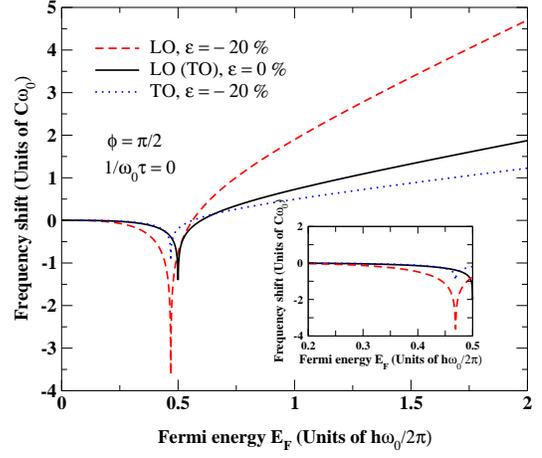}
\end{center}
\caption{Frequency shifts of LO (dashed line) and TO (dotted line) modes as a function of the Fermi energy $E_F$ in the clean 
limit ($\frac 1{\tau\omega_0}=0$) and for a compressive strain $\epsilon=\frac{\delta a}a=-20\%$. The LO mode is along the strain axis.
The solid line is the result for the undeformed case. The inset shows the frequency shifts for $E_F< \frac{\hbar \omega_0}2$.}
\label{shift0}
\end{figure}

For clarity reasons, we will consider in the following strain strength $|\epsilon| \geq 10\%$. It should be noted that the critical 
strain for graphene is of $25\%$.\newline
Figure \ref{shift0} shows the dependence of the frequency shifts on the Fermi energy $E_F$ in the clean limit ($\frac 1{\tau\omega_0}=0$) 
for a compressive strain strength $\epsilon=-20\%$. \\

Due to the deformation, the degeneracy of LO and TO modes, obtained in the undeformed graphene (solid line in Fig.\ref{shift0}), is lifted.\newline
The logarithmic singularity at $E_F=\frac{\hbar \omega_0}2$ reported in the undeformed case is a robust feature which persists under strain 
but takes place at $E_F= \frac{\hbar \omega_0}2 \left( 1+\frac{\epsilon}3\right)$ which corresponds to $E^{\ast}_F=\frac{\hbar \omega_0}2$ 
in Eq.\ref{self5}.\

%%%%%%%%Fig.
\begin{figure}[hpbt] 
\begin{center}
\vspace{0.5cm}
\includegraphics[width=0.8\columnwidth]{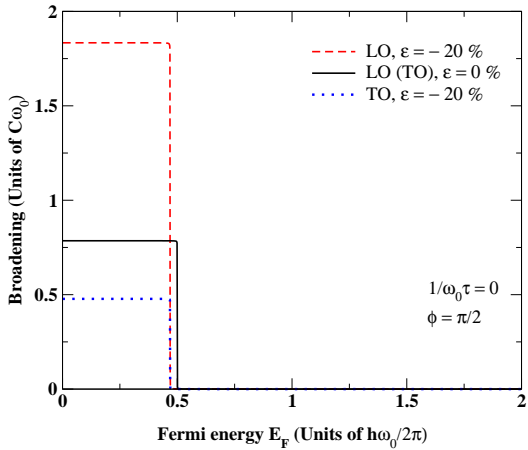}
\end{center}
\caption{Broadenings of LO (dashed line) and TO (dotted line) modes as a function of the Fermi energy $E_F$ in the clean limit ($\frac 1{\tau\omega_0}=0$) 
and for a compressive strain $\epsilon=\frac{\delta a}a=-20\%$. The LO mode is along the strain axis.
The solid line is the result for the undeformed case.}
\label{broad0}
\end{figure}

According to Fig.\ref{shift0},  both TO and LO modes are redshifted leading to a lattice softening for $E_F<\frac{\hbar \omega_0}2 
\left( 1+\frac{\epsilon}3\right)$. However, the phonon frequencies increase with $E_F$ and the lattice hardens for 
$E_F>\frac{\hbar \omega_0}2 \left( 1+\frac{\epsilon}3\right)$.
Moreover, the frequency of the LO mode, which is along the strain axis, is more shifted compared the the TO mode. The LO mode is, 
then, more affected by the electron-phonon interaction as shown by the broadening behavior depicted in figure \ref{broad0}.
The damping of the LO mode is more pronounced than that of the TO mode which is found to be more long lived than the modes of undeformed graphene.\

This behavior can be understood from the structure of the electronic dispersion.
Along the strain direction, the electron velocity is enhanced for a compressive deformation ($\epsilon<0$) 
as $v_y=\frac{w_y}{\hbar}\simeq \frac 3{2\hbar} (1-\frac 43 \epsilon)at$, while that in the perpendicular direction 
is reduced as $v_x=\frac{w_x}{\hbar}\simeq \frac 3{2\hbar} (1+\frac 23 \epsilon)at$. \

\begin{figure}[hpbt] 
\begin{center}
\includegraphics[width=0.6\columnwidth]{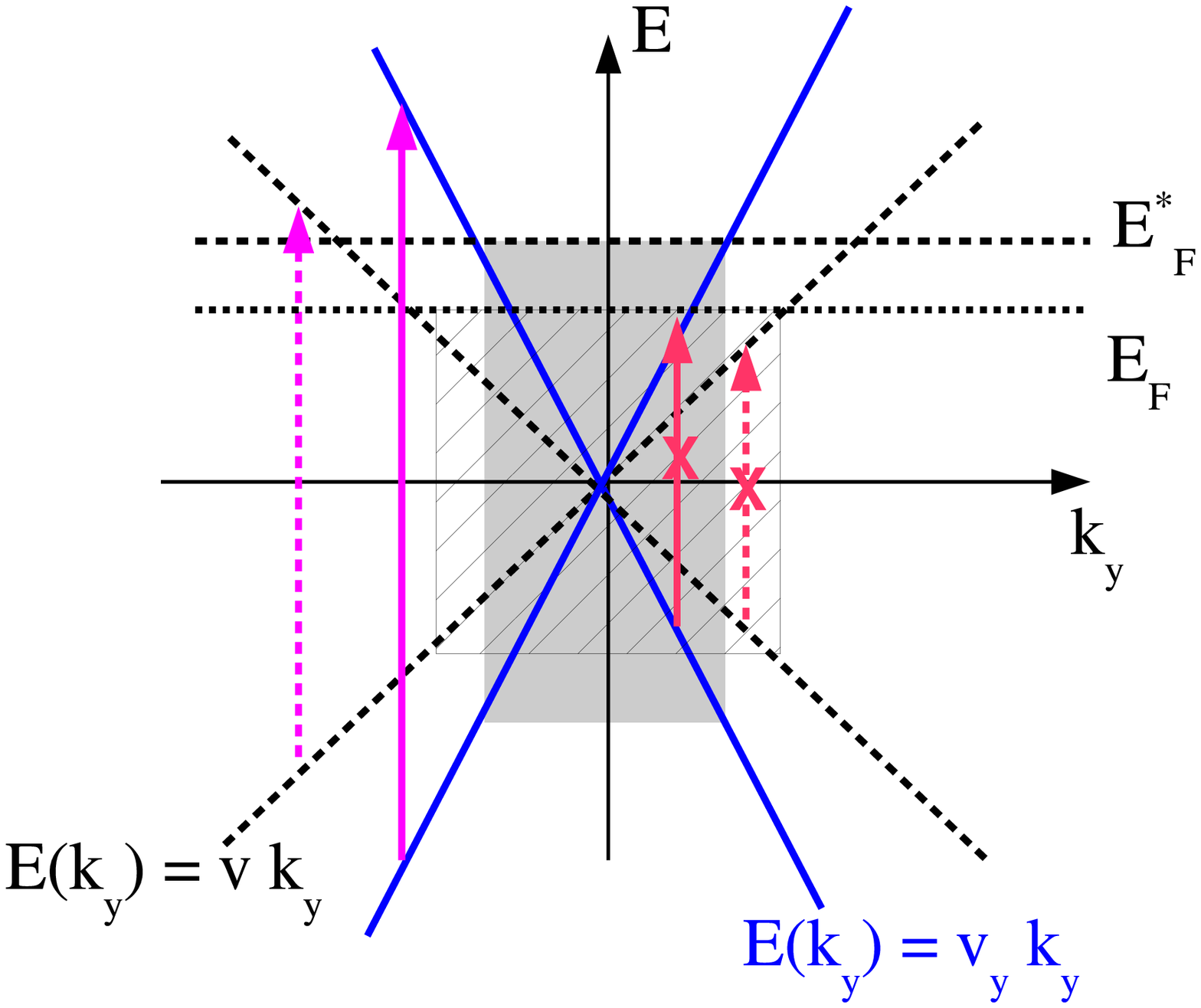}
\includegraphics[width=0.6\columnwidth]{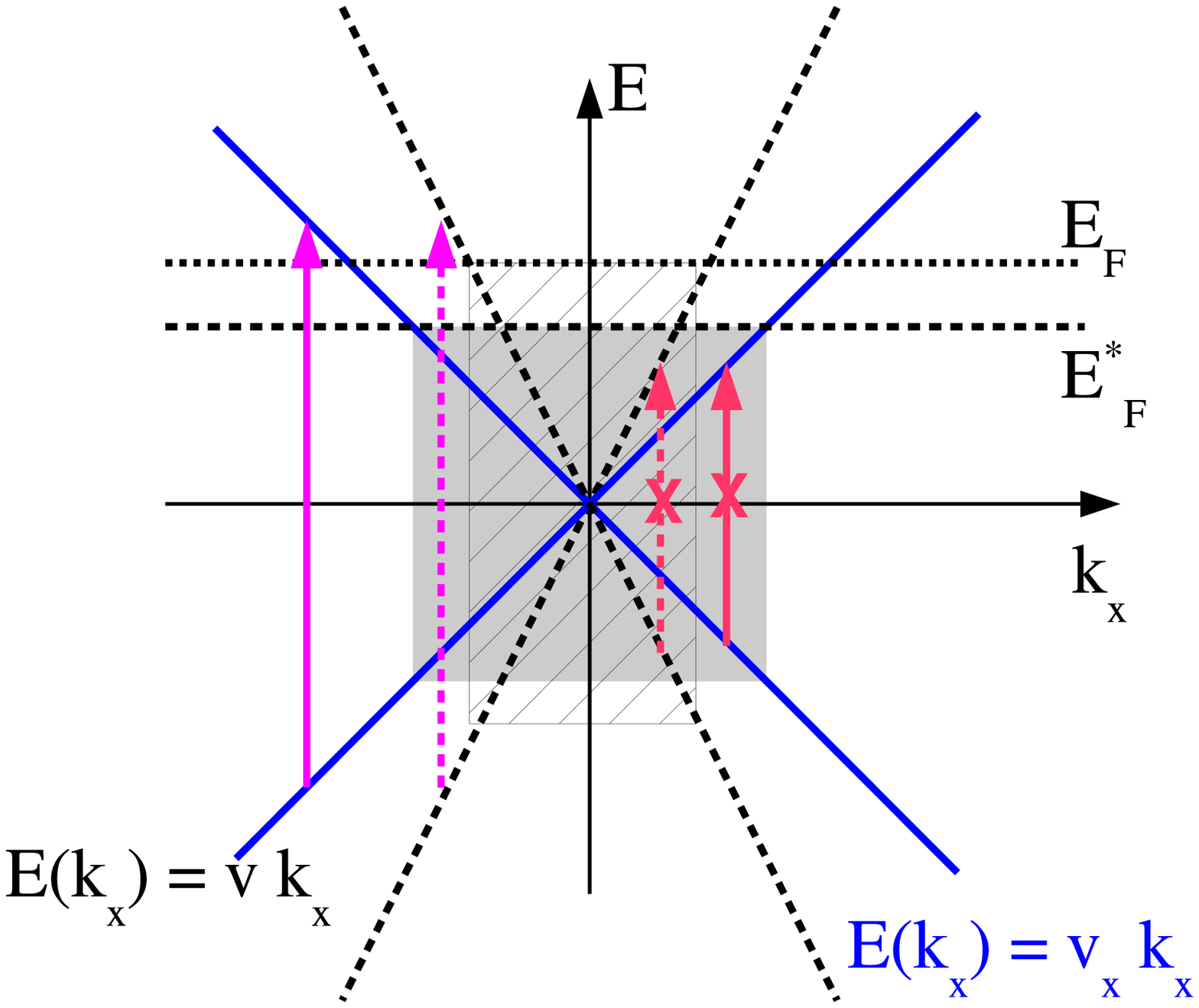}
\end{center}
\caption{Electron-hole process responsible of phonon hardening corresponds to the states where production of electron-hole pairs is 
forbidden by Pauli principle. These states correspond to the dashed region for undeformed case and grey area for compressive strain.
The Fermi level ($E^{\ast}_F$) increases under compressive deformation and the Fermi velocity $v_y$ ($v_x$) along (perpendicular) to 
the strain direction $(y^{\prime}y$) is enhanced (reduced) compared to isotropic case. This leads to more (less) electron-hole pairs 
contributing to phonon softening. The dashed and solid arrows (crossed solid and dashed arrows) denote the electron-hole process leading 
to phonon softening (hardening) for undeformed and compressed case respectively.}
\label{pauli}
\end{figure}

The Fermi level changes as $E^{\ast}_F\simeq E_F\left(1-\frac{\epsilon}3\right)$ which increases for a compressive strain (Fig.\ref{pauli}).
As a consequence, the production of electron-hole pairs is furthered along the strain direction, as shown in figure \ref{pauli}, 
since there are more states which are not blocked by Pauli principle for a given phonon frequency.
However, in the direction perpendicular to the strain, electron-hole processes, allowed in the undeformed case, become forbidden 
by the Pauli exclusion principle.
This explains the long lived TO phonon mode compared to the modes of undeformed graphene.\\

The behavior of LO and TO modes are exchanged for $\Phi=0$ where the TO mode becomes along the strain direction. Moreover, the behavior 
are also exchanged for tensile deformation ($\epsilon>0$).\\

This feature can be understood from Eq.\ref{self5} showing that the leading term for the frequency shifts is $\frac1{\alpha^{\prime 2}}>1$ 
in compressive strain and $\alpha^{\prime 2}>1$ for tensile deformation.\\

\begin{figure}[hpbt] 
\begin{center}
\vspace{0.5cm}
\includegraphics[width=0.8\columnwidth]{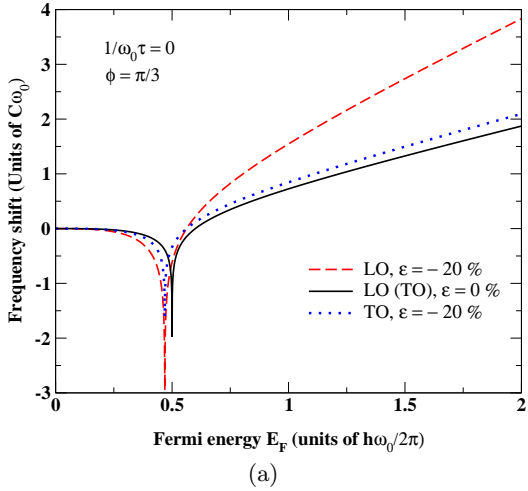}\\

(a)
\vspace{1cm}

\includegraphics[width=0.8\columnwidth]{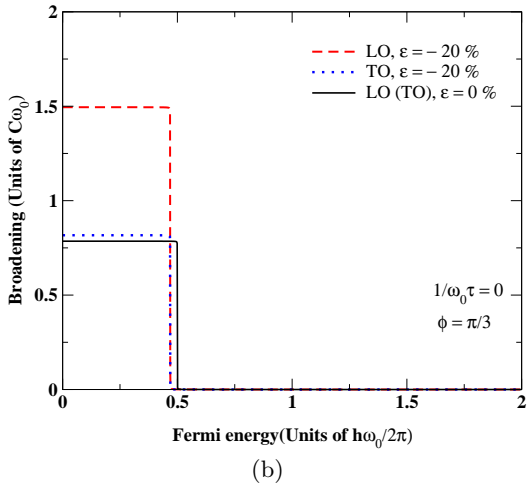}\\

(b)
\end{center}
\caption{Frequency shifts and broadenings of LO (dashed line) and TO (dotted line) modes as a function of the Fermi energy $E_F$ in 
the clean limit ($\frac 1{\tau\omega_0}=0$) and for a compressive deformation $\epsilon=\frac{\delta a}a=-20\%$. The phonon angle is $\Phi=\frac{\pi} 3$.
The solid line is the result for the undeformed case.}
\label{shiftphi}
\end{figure}

Figure \ref{shiftphi} shows the frequency shifts and the broadening of the phonon modes at $\Phi=\frac{\pi} 3$. The difference in damping 
of TO and LO modes, obtained for $\Phi=\frac{\pi }2$ and $\Phi=0$, is clearly reduced since both modes have a component along the strain direction.\

\begin{figure}[hpbt] 
\begin{center}
\vspace{0.5cm}
\includegraphics[width=0.8\columnwidth]{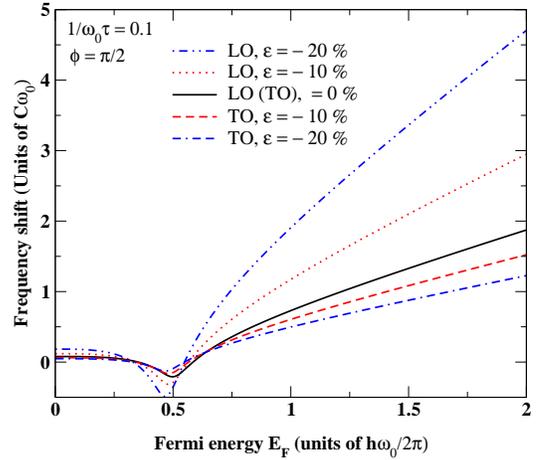}
\end{center}
\caption{Frequency shifts of LO and TO modes as a function of the Fermi energy $E_F$ in the dirty limit ($\frac 1{\tau\omega_0}=0.1$) 
and for compressive strains of $-10\%$ and $-20\%$. The LO phonon mode is along the strain direction.
The solid line is the result for the undeformed case.}
\label{loga}
\end{figure}

The logarithmic singularity obtained in the clean limit at $E^{\ast}_F=\frac{\hbar \omega_0}2$ (Fig.\ref{shift0}) is smeared out in 
the dirty limit as shown in Fig.\ref{loga} for $\Phi=\frac{\pi }2$ in the case of tensile and compressive deformation.
According to Fig.\ref{loga}, the frequency shifts of LO and TO modes depend on the Fermi level and the amount of disorder.
Away from $E^{\ast}_F\sim\frac{\hbar \omega_0}2$, all modes show a blueshift contrary to the clean limit where LO and 
TO modes undergo a redshift (blueshift) for compressive (tensile) strain at $E^{\ast}_F<\frac{\hbar \omega_0}2$. 
The frequency blueshift is reminiscent of that found by Ando\cite{ando2006} in undeformed graphene in the dirty limit.\\

The dependence of the frequency shifts on the doping level and the amount of disorder may explain the discrepancy in 
the experimental values of the shift rates of G$^+$ and G$^-$ bands as function of the strain \cite{MHuang, Ninano, Mohiuddin,frank2,C-huang2013} 
and which was ascribed to a difference in the strain calibration. We suggest that, this discrepancy may be due to the doping and the disorder 
amount in the sample. \\

In Ref.\cite{C-huang2013}, the authors studied the behavior of the G band in deformed graphene using polarized light. They reported that the 
G peak can be regarded as mixture of three peaks corresponding to undeformed case (G$^0$), compressive (G$^-$) and tensile (G$^+$) deformation. 
The authors attributed the presence of both blue and red shifted frequencies (G$^+$ and G$^-$ bands) to the anisotropy of the applied deformation.
According to Figs.\ref{shift0} and \ref{loga}, for $E^{\ast}_F>\frac{\hbar \omega_0}2$ and $\Phi=\frac{\pi }2$, the LO mode (TO mode) 
is blueshifted (redshifted) compared to the undeformed mode (solid line in the figures) for compressive strain.
The experimental results of Ref.\cite{C-huang2013} could then be the signature of the electron-phonon interaction. 
The shifted G$^+$ and G$^-$ modes could be assigned to the LO and TO modes for a given uniaxial strain at a doping 
level $E^{\ast}_F>\frac{\hbar \omega_0}2$.\\

\begin{figure}[hpbt] 
\begin{center}
\vspace{0.5cm}
\includegraphics[width=0.8\columnwidth]{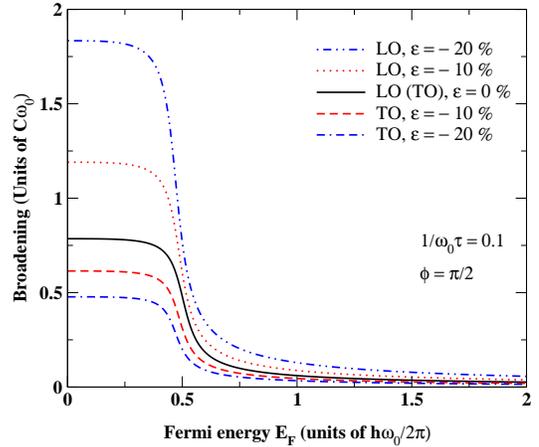}
\end{center}
\caption{Broadenings of LO and TO modes as a function of the Fermi energy $E_F$ in the dirty limit ($\frac 1{\tau\omega_0}=0.1$) 
and for compressive strains of $-10\%$ and $-20\%$. The LO phonon mode is along the strain direction. 
The solid line is the result for the undeformed case.}
\label{broad}
\end{figure}

In figure \ref{broad}, we plot the broadening of phonon modes as a function of the Fermi energy for $\Phi=\frac{\pi }2$ in the dirty limit. 
The figure shows that the damping of the mode along the strain direction is enhanced as the amplitude of the deformation increases. 
This reflects the increasing number of the electron-hole pairs leading to decaying phonons (Fig.\ref{pauli}). \

\begin{figure}[hpbt]
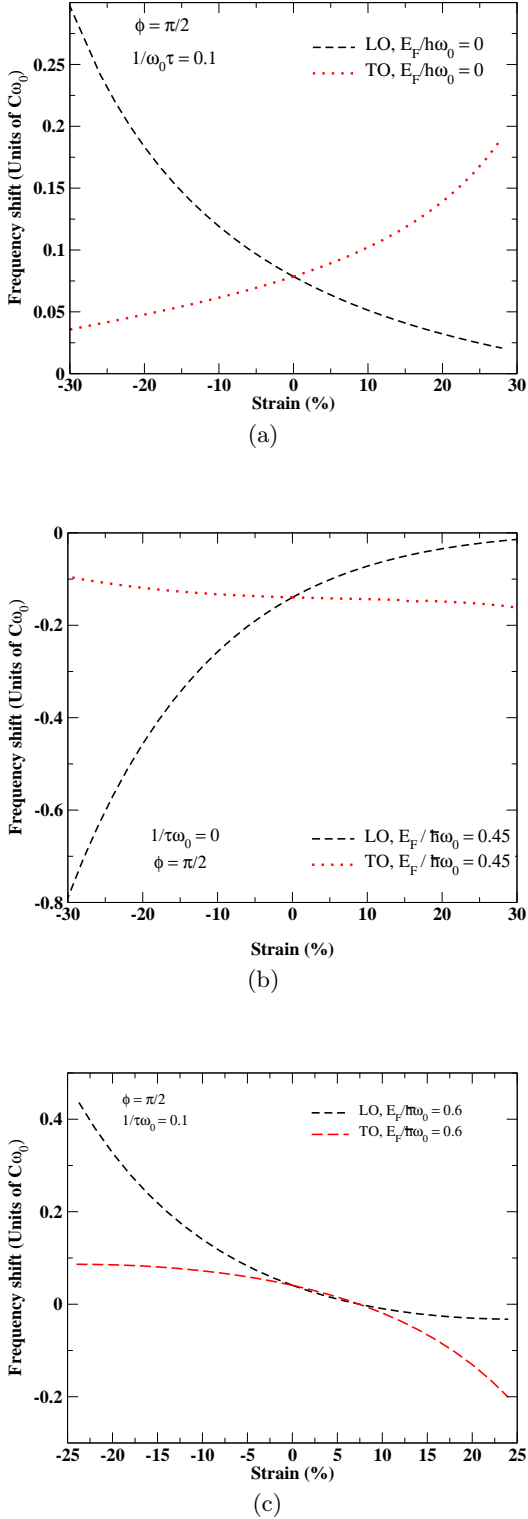
 
\begin{center}
\vspace{0.5cm}
\includegraphics[width=0.8\columnwidth]{shiftstrain.eps}\\

(a)
\vspace{1cm}

\includegraphics[width=0.8\columnwidth]{shiftstrain2.eps}\\

(b)
\vspace{1cm}

\includegraphics[width=0.8\columnwidth]{shiftstrain06.eps}\\

(c)
\end{center}
\caption{Strain dependence of the of LO (dashed line)  and TO (dotted line) frequency shifts in (a) undoped case and in the 
dirty limit ($\frac 1{\tau\omega_0}=0.1$), (b) in doped case ($\frac{E_F}{\hbar\omega_0} =0.45$) and in the clean 
limit ($\frac 1{\tau\omega_0}=0$) and (c) in doped case ($\frac{E_F}{\hbar\omega_0} =0.6$) and in the dirty 
limit ($\frac 1{\tau\omega_0}=0.1$. The LO phonon mode is along the strain direction.}
\label{shiftstrain}
\end{figure}

The strain dependence of the frequency shifts is depicted in Fig.\ref{shiftstrain} where we considered the case of undoped graphene 
in the dirty limit ($E_F=0$, $\frac1{\tau\omega_0}=0.1$) and the doped graphene ($\frac{E_F}{\hbar\omega_0} =0.45$) 
in the clean limit since the shifts in the clean undoped case are small. The shift behaviors could be understood from the processes depicted 
in Fig.\ref{pauli}.\

Fig.\ref{shiftstrain} shows a linear behavior of the frequency shift as a function of the strain strength for small strain. 
This is reminiscent of the experimental results reported in Refs. \cite{Mohiuddin,frank2}. The strain rates and slopes of 
the frequency shifts are dependent on the doping level and the disorder amount.\

According Fig.\ref{shiftstrain}, the linearity is lost by increasing the strain. It is worth to note that a departure from 
a linear behavior was also reported in Ref.\cite{Son2011} for the strain dependence of the frequency shift of the 2D Raman band.
Such behavior could also be observed in Raman spectra of $\alpha$(BEDT)$_2$I$_2$ salt showing a strong anisotropic electronic Dirac spectrum. \\

\begin{figure}[hpbt] 
\begin{center}
\vspace{0.5cm}
\includegraphics[width=0.8\columnwidth]{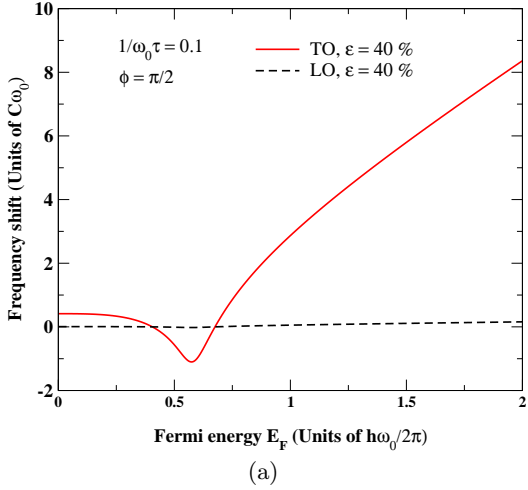}\\

(a)
\vspace{1cm}

\includegraphics[width=0.8\columnwidth]{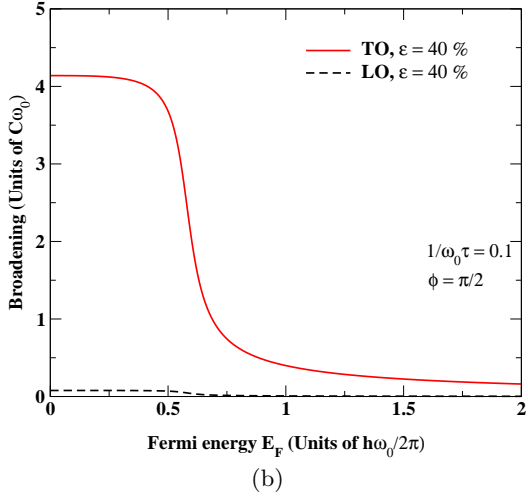}\\

(b)
\end{center}
\caption{(a) Frequency shifts and (b) broadenings of the of LO (dashed line) and TO (solid line) as a function of 
the Fermi energy $E_F$ in the dirty limit ($\frac 1{\tau\omega_0}=0.1$) and for a strong tensile deformation. 
The LO phonon mode, which is along the strain direction, decouples from the electron-hole pairs.}
\label{shift4}
\end{figure}

In the limit of strong strain, we expect a decoupling of electron-hole pairs from the phonon mode along (perpendicular) 
to the strain axis for tensile (compressive) deformation as shown in Fig.\ref{shift4}. Such effect could not be observed 
in graphene where the critical strain is of 25$\%$ but may be bring out in $\alpha$(BEDT)I$_2$ \cite{Frederic,Pasquier}.\\

A hallmark feature of the doping dependence of the frequency shifts is the presence of crossings of LO and TO modes (Figs.\ref{shift0}, \ref{loga}). 
At the corresponding Fermi energy, no G band splitting is expected due to electron-phonon interaction. Experimentally, the G$^+$ and the G$^-$ 
bands should then merge in uniaxial strained graphene by doping the sample at the critical value corresponding to the crossing of LO and TO modes. 
This feature could only be observed in the absence of the shear strain which induces a splitting of the G band.
A possible crossing of LO and TO modes was also reported in carbon nanotubes \cite{Sasaki2008}.\\

\begin{figure}[hpbt] 
\begin{center}
\vspace{0.5cm}
\includegraphics[width=0.8\columnwidth]{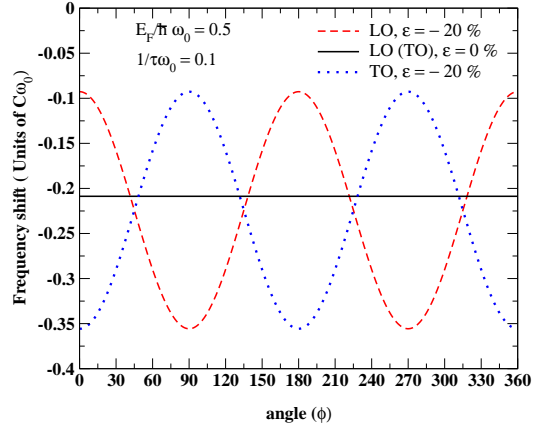}
\end{center}
\caption{Angle dependence of the frequency shifts of the LO (dashed line) and TO (dashed-dotted line) in the doped case ($E_F=0.5\hbar\omega_0$), 
in the dirty limit ($\frac 1{\tau\omega_0}=0.1$) and for a compressive deformation. $\Phi$ (in degree) is the angle of optical phonon with 
respect to $x$ axis perpendicular to the strain direction. The solid line is the result for the undeformed case}
\label{phi}
\end{figure}

In figure \ref{phi}, we plot the dependence of the phonon frequency shifts on the phonon angle $\Phi$ with respect to the axis perpendicular 
to the strain direction. The shifts of the LO and TO modes display a periodic modulation with a relative shift of 90$^{\circ}$. 
According to Eq.\ref{self5}, this dependence is due to the anisotropy of the electronic dispersion relation. 
Considering the isotropic case ($\alpha^{\prime}$=1), the shifts become independent on $\Phi$ as in isotropic honeycomb lattice \cite{ando2006}.\

Our results are in agreement with the experimental data \cite{Mohiuddin,C-huang2013} and numerical calculations \cite{Popov} 
showing a periodic modulation of the intensity of G$^+$ and the G$^-$ peaks as a function of the angle between the incident 
light polarization and the strain axis. The relative shifts of the two bands is also found to be of 90$^{\circ}$.
Our results support the idea presented in the experimental study of Mohiuddin \textit{et al.}\cite{Mohiuddin} suggesting 
that the polarization dependence of the G peaks is due to the anisotropy of the electronic spectrum and such dependence is 
the signature of the electron-phonon intercation. 
It is worth to stress that Sasaki {\it et al.}\cite{Sasakiedge} proposed that the nature of the graphene edges contributes 
also to the polarization dependence of Raman bands in strained graphene.

%%%%%%%%%%%%%
\section{Concluding remarks}
We have derived the frequency shifts and the broadenings of the longitudinal (LO) and transverse (TO) optical phonon modes at $\Gamma$ 
point in graphene under uniaxial strain disregarding the contribution of shear strain component. We show that the Raman G band, 
corresponding to a double degenerate mode in undeformed graphene, may split into two peaks due to electron-phonon interaction. 
These peaks are assigned to the LO and the TO modes which are found to be strongly dependent on the Fermi level and the amount of disorder. 
This dependence may explain the difference in the experimental results giving the strain rates of the frequency shifts of the G$^+$ and G$^-$ modes.\

Moreover, we found that the splitting of the G band is due to the anisotropy of the electronic spectrum. The tilt of Dirac cones, arising also 
from the strain, is found to be irrelevant for the relative frequency shift of the LO and TO modes since it leads to a global shift of the G peak. \

We also show that the electron-phonon intercation contributes to the Raman polarization dependence of the G peaks in strained graphene. 
This contribution reflects the anisotropy of the electronic spectrum. The optical phonon mode along the strain is found to be damped 
(long lived) for compressive (tensile) strain. The frequency shifts and the lifetime of the optical phonons are substantially dependent 
on the strain strength and the phonon angle. At relatively strong strain, it is possible to induce a decoupling of the phonon mode perpendicular 
to the compressive strain axis from electron-hole pair production process.
The signature of the strain induced anisotropic electronic dispersion could also be brought out in the $\Gamma$ point magnetophonon 
resonance at high magnetic field \cite{Assili2}.

\section{Acknowledgment}

We warmly thank Y.-W. Son, K. Sasaki for helpful and stimulating discussions.
We thank J.-N. Fuchs for a critical reading of the manuscript.
This work was partially supported by the National Research Foundation of Korea (NRF) grant funded by the Korea government (MEST) (No. 2011-0030902).
S. H acknowledges the kind hospitality of W. Kang and the members of CERC (Seoul, Korea). The final form of the manuscript was prepared in 
the ICTP (Trieste, Italy). S. H. was supported by Simons-ICTP associate fellowship.

%%%%%%%%%%%%%%%%%%%% References %%%%%%%%%%%%%%%%%%%%%%
\appendix
%\begin{appendix}
\section{Weyl Hamiltonian by $\vec{k}.\vec{p}$ method}

The $\vec{k}.\vec{p}$ method was used by Ando \cite{revando,ando2006} to derive the Dirac Hamiltonian in undeformed graphene taking only 
into account the hopping integral to the first neighboring carbon atoms. Following Ref.\cite{ando2006}, we derive the Weyl Hamiltonian 
for uniaxial strained graphene considering first and second neighboring hopping integrals.
We start with the eigenproblem given by Eq.\ref{eigen} where the functions $\psi_A(\vec{R}_A)$ and $\psi_B(\vec{R}_B)$ are written as:
\begin{eqnarray}
\psi_A(\vec{R}_A)=a^{\dagger}(\vec{R}_A)\Phi_A(\vec{R}_A)\nonumber\\
\psi_B(\vec{R}_B)=b^{\dagger}(\vec{R}_B)\Phi_B(\vec{R}_B)\nonumber\\
\end{eqnarray}
here the vectors $a(\vec{R}_A)$, $b(\vec{R}_B)$, $\Phi_A(\vec{R}_A)$ and $\Phi_B(\vec{R}_B)$ are given by:
\begin{eqnarray}
a(\vec{R}_A)=\left(
\begin{array}{c}
\mathrm{e}^{-i\vec{k}^D.\vec{R}_A}\\
\mathrm{e}^{-i\vec{k}^{D^{\prime}}.\vec{R}_A}
\end{array}
\right)\quad
b(\vec{R}_B)=\left(
\begin{array}{c}
\mathrm{e}^{-i\vec{k}^D.\vec{R}_B}\\
-\mathrm{e}^{-i\vec{k}^{D^{\prime}}.\vec{R}_B}
\end{array}
\right)\nonumber\\
\Phi_A(\vec{R}_A)=\left(
\begin{array}{c}
F_A^D(\vec{R}_A)\\
F_A^{D^{\prime}}(\vec{R}_A)
\end{array}
\right)\quad
\Phi_B(\vec{R}_B)=\left(
\begin{array}{c}
F_B^D(\vec{R}_B)\\
F_B^{D^{\prime}}(\vec{R}_B)
\end{array}\right)\nonumber\\
\end{eqnarray}
The l.h.s of Eq.\ref{eigen} can be written, at $\vec{R}_A$ as:
\begin{eqnarray}
 \varepsilon a^{\dagger}(\vec{R}_A)F_A(\vec{R}_A)
=\varepsilon \sum_{\vec{R}_A} g(\vec{r}-\vec{R}_A)a(\vec{R}_A)a^{\dagger}(\vec{R}_A)F_A(\vec{R}_A)\nonumber\\
\end{eqnarray}
where $g(\vec{r})$ is a smoothing function satisfying:
\begin{eqnarray}
\sum_{\vec{R}_A} g(\vec{r}-\vec{R}_A)=\sum_{\vec{R}_B} g(\vec{r}-\vec{R}_B)=1\nonumber\\
f(\vec{r})g(\vec{r}-\vec{R}_A)\simeq f(\vec{R})g(\vec{r}-\vec{R}).
\end{eqnarray}
$f(\vec{r})$ is an envelope function\cite{ando2006}.
These properties yield to
\begin{widetext}
 \begin{eqnarray}
\sum_{\vec{R}_A} g(\vec{r}-\vec{R}_A)\mathrm{e}^{i(\vec{k}^{D^{\prime}}-\vec{k}^{D}).\vec{R}_A}=
\sum_{\vec{R}_B}
g(\vec{r}-\vec{R}_B)\mathrm{e}^{i(\vec{k}^{D^{\prime}}-\vec{k}^{D}).\vec{R}_B}\simeq 0\nonumber\\
\sum_{\vec{R}_A} g(\vec{r}-\vec{R}_A)a(\vec{R}_A)a^{\dagger}(\vec{R}_A)\simeq \left(
\begin{array}{cc}
1&0\\
0&1
\end{array}\right)
\end{eqnarray} 
\end{widetext}
which is reminiscent of the $\delta$ function \cite{ando2006}.
Eq.\ref{eigen} can then be written, around the A site, as:
\begin{widetext}
\begin{eqnarray}
\varepsilon \sum_{\vec{R}_A}g(\vec{r}-\vec{R}_A)a(\vec{R}_A)a^{\dagger}(\vec{R}_A)F_A(\vec{r})=
&-&\sum_{l=1}^3 t_{nn}^{(l)}\sum_{\vec{R}_A} g(\vec{r}-\vec{R}_A)a(\vec{R}_A)b^{\dagger}(\vec{R}_B)F_B(\vec{r}-\vec{\tau}_l)\nonumber\\
&-&\sum_{l=1}^6 t_{nnn}^{(l)}\sum_{\vec{R}_A} 
g(\vec{r}-\vec{R}_A)a(\vec{R}_A)a^{\dagger}(\vec{R}_A-\vec{a}_l)F_A(\vec{r}-\vec{a}_l)
\label{eigen2}
\end{eqnarray} 
\end{widetext}

The l.h.s of Eq.\ref{eigen2} reduces to $\varepsilon F_A(\vec{R}_A)$ and, in the r.h.s, we set:
\begin{eqnarray}
F_B(\vec{r}-\vec{\tau}_l)&&\simeq F_B(\vec{r})-\left(\vec{\tau}_l.\frac{\partial}{\partial \vec{r}}\right)F_B(\vec{r})\nonumber\\
F_A(\vec{r}-\vec{a}_l)&&\simeq F_A(\vec{r})-\left(\vec{a}_l.\frac{\partial}{\partial \vec{r}}\right)F_A(\vec{r})\nonumber\\
\label{dl}
\end{eqnarray}
We then obtain
\begin{eqnarray}
\sum_{\vec{R}_A}g(\vec{r}-\vec{R}_A)a(\vec{R}_A)b^{\dagger}(\vec{R}_A-\vec{\tau}_l)\simeq
\left(
\begin{array}{cc}
\mathrm{e}^{i\vec{k}^{D}.\vec{\tau}_l}&0\\
0&-\mathrm{e}^{-i\vec{k}^{D^{\prime}}.\vec{\tau}_l}
\end{array}
\right)\nonumber\\
\end{eqnarray} 
Applying this term to $F_B(\vec{r})$ in Eq.\ref{eigen2} and summing over $l$ gives rise to a diagonal term of the 
form $(2\cos \theta)1\!\!1=-\frac{t^{\prime}}{t}1\!\!1$ which leads to a shift of the total energy.\

The term $\left(\vec{\tau}_l.\frac{\partial}{\partial \vec{r}}\right)F_B(\vec{r})$ in Eq.\ref{dl}, summed over $l$ and applied to $F_B(\vec{r})$, gives:
\begin{widetext}
\begin{eqnarray}
\sum_l\left(\vec{\tau}_l.\frac{\partial}{\partial \vec{r}}\right)
\left(
\begin{array}{c}
F_B^D(\vec{r})\\
F_B^{D^{\prime}}(\vec{r})
\end{array}
\right)
=\left(
\begin{array}{c}
t\left(-a\sqrt{3}\sin \theta\, k_x+ia\cos\theta\, k_y-iat^{\prime}(1+\epsilon)k_y\right)F_B^D(\vec{r})\\
t\left(a\sqrt{3}\sin\theta\, k_x+ia\cos\theta\, k_y-iat^{\prime}(1+\epsilon)k_y\right)
F_B^{D^{\prime}}(\vec{r})
\end{array}
\right)
\end{eqnarray}  
\end{widetext}
In Eq.\ref{eigen2}, the contribution of the first neighbor hopping integrals gives then rise to the following eigenproblem near $D$ point:
\begin{eqnarray}
 \varepsilon F_A(\vec{r})=\left(
\begin{array}{cc}
 w_xk_x-iw_yk_y&0\\
0&w_xk_x+iw_yk_y
\end{array}
\right)F_B(\vec{r})\nonumber\\
\end{eqnarray}
where $w_x=\sqrt{3}a t \sin \theta$, $w_y=-ta\cos\theta+t^{\prime}a(1+\epsilon)=
\frac 32 t^{\prime}(1+\frac 23 \epsilon)a$, $k_x=-i\frac {\partial}{\partial x}$ and $k_y=-i\frac {\partial}{\partial y}$.\\

For the second neighbor hopping integrals, one have:
\begin{eqnarray}
\sum_{\vec{R}_A}g(\vec{r}-\vec{R}_A)a(\vec{R}_A)a^{\dagger}(\vec{R}_A-\vec{a}_l)\simeq
\left(
\begin{array}{cc}
\mathrm{e}^{i\vec{k}^{D}.\vec{a}_l}&0\\
0&\mathrm{e}^{-i\vec{k}^{D^{\prime}}.\vec{a}_l}
\end{array}
\right)\nonumber\\
\end{eqnarray} 
and 
\begin{eqnarray}
\sum_l t_{nnn}^{(l)} \mathrm{e}^{-i\vec{k}^{D}.\vec{a}_l} \left(\vec{a}_l.\frac{\partial}{\partial \vec{r}}\right)=w_{0x}k_x
\end{eqnarray}
where $w_{0x}=2\sqrt{3}a(t_{nnn}\sin 2\theta +t_{nnn}^{\prime}\sin \theta)$.\\

The electronic Hamiltonian, near $D$ and $D^{\prime}$ points, takes the form:
\begin{eqnarray}
 H^D_{\xi}=\xi\left(
\begin{array}{cc}
 w_{0x}k_x& w_xk_x-i\xi w_yk_y\\
w_xk_x+i\xi w_yk_y& w_{0x}k_x
\end{array}
\right)
\label{weyl}
\end{eqnarray}
with $\xi=+$ (-) at $D$ ($D^{\prime}$) point.\

$w_x$ and $w_y$ can be expressed as a function of the strain strength as
\begin{eqnarray}
w_x=\sqrt{3}a t \sin \theta\simeq \frac 32 at\left(1+\frac 23 \epsilon\right)\\
w_y=\frac 32 t^{\prime}(1+\frac 23 \epsilon)a\simeq \frac 32 at \left(1-\frac 43 \epsilon\right)
\end{eqnarray}
In graphene, $\tilde{w}_0=\sqrt{\left(\frac{w_{0x}}{w_x}\right)^2+ \left(\frac{w_{0y}}{w_y}\right)^2}\simeq 0.6\epsilon$ \cite{mark2008}. 
In the present case, we have $w_{0y}=0$.\\
%\end{appendix}

%\begin{appendix}
\section{Electron-phonon effective Hamiltonian}
Regarding the effect of the lattice distortion on the hopping integral (Eq.\ref{tnn})
an extra term appears in the electronic Hamiltonian (Eq.\ref{weyl}). This term arises from the contribution of the hopping 
term correction $\frac{\partial t_{nn}^{(l)}}{\partial d_l}$ in Eq.\ref{eigen}. This contribution is of the form
\begin{widetext}
\begin{small}
 \begin{eqnarray}
\sum_l\sum_{\vec{R}_A}g(\vec{r}-\vec{R}_A)a(\vec{R}_A)b(\vec{R}_A-\vec{\tau}_l)
\left(-\frac{\partial t_{nn}^{(l)}}{\partial d_l}\right)\sqrt{2}\left(\frac{\vec{\tau}_l}{d_l}\right).\vec{u}(\vec{r})F_B(\vec{r})
=\sum_l\left(
\begin{array}{cc}
\mathrm{e}^{-i\vec{k}^{D}.\vec{\tau}_l}&0\\
0&-\mathrm{e}^{-i\vec{k}^{D^{\prime}}.\vec{\tau}_l}
\end{array}
\right)
\left(-\frac{\partial t_{nn}^{(l)}}{\partial d_l}\right)\sqrt{2}\left(\frac{\vec{\tau}_l}{d_l}\right).\vec{u}(\vec{r})F_B(\vec{r})\nonumber\\
\end{eqnarray} 
\end{small}
\end{widetext}
where the summation over $l$ around $D$ point gives:
\begin{widetext}
\begin{eqnarray}
\sum_l -\mathrm{e}^{-i\vec{k}^{D}.\vec{\tau}_l}\left(-\frac{\partial t_{nn}^{(l)}}{\partial d_l}\right)\sqrt{2}
\left(\frac{\vec{\tau}_l}{d_l}\right).\vec{u}(\vec{r})=\frac{\sqrt{2}}{ta}
\left(\frac{\partial t}{\partial a}\right)\left[iw_xu_x+w_y^{\prime}u_y\right]
\end{eqnarray} 
\end{widetext}
where $d_l=a$ and we used the Harrison's law \cite{mark2008}:
$\frac 1{t_{nn}d_l}\left(\frac{\partial t_{nn}^{(l)}}{\partial d_l}\right)=-\frac{2}{d_l^2}$.
Here $w_y^{\prime}=w_y-2\epsilon t^{\prime}a(1+\epsilon)\simeq w_y(1-\frac 43 \epsilon)$.\

This contribution gives rise to the effective phonon-electron Hamiltonian given by Eq.\ref{DeltaH}.
%\end{appendix}
%%%%

\end{document}